\pdfoutput=1

\documentclass[twocolumn,english,superscriptaddress]{revtex4}
\usepackage{times,amssymb,amsmath,graphicx}
\usepackage[bookmarks=false,pdffitwindow=false,pdfstartview={FitH}]{hyperref}
\usepackage[T1]{fontenc}
\usepackage{babel}
\usepackage{color}

\begin{document}


\title{
  Sub-Sharvin conductance and incoherent shot-noise in 
  graphene disks at magnetic field
}

\author{Adam Rycerz\footnote{Corresponding author; e-mail:
  \href{mailto:rycerz@th.if.uj.edu.pl}{rycerz@th.if.uj.edu.pl}.}}
\affiliation{Institute for Theoretical Physics,
  Jagiellonian University, \L{}ojasiewicza 11, PL--30348 Krak\'{o}w, Poland}

\author{Katarzyna Rycerz}
\affiliation{Institute of Computer Science, AGH University of Science
  and Technology, al.\ Mickiewicza 30, PL--30059 Krak\'{o}w, Poland}

\author{Piotr Witkowski}
\affiliation{Institute for Theoretical Physics,
  Jagiellonian University, \L{}ojasiewicza 11, PL--30348 Krak\'{o}w, Poland}

\date{April 19, 2024}

\begin{abstract}
  Highly-doped graphene samples show the conductance reduced and the
  shot-noise power enhanced compared to standard ballistic systems in
  two-dimensional electron gas. These features can be understood within
  a~model assuming incoherent scattering of Dirac electrons between two
  interfaces separating the sample and the leads. 
  Here we find, by adopting the above-mentioned model for the edge-free
  (Corbino) geometry and by means of the computer simulation of quantum
  transport, that another graphene-specific feature should be observable
  when the current flow through a~doped disk is blocked by high magnetic
  field. In case the conductance drops to zero, the Fano factor approaches
  the value of $F\approx{}0.56$, with a~very weak dependence on the disk
  radii ratio. The role of finite source-drain voltages and the system
  behavior upon tuning the electrostatic potential barrier from a rectangular
  to parabolic shape are also discussed. 
\end{abstract}

\maketitle


\section{Introduction}
Although electronic properties of matter are governed by the rules of
quantum mechanics \cite{Dat05}, it is very unlikely to find that any
measurable characteristic of a~macroscopic system is determined solely by
the universal constants of nature, such as the elementary charge ($e$) or
the Planck constant ($h$).
In the last century, two notable exceptions arrived with the phenomena
of superconductivity \cite{Kit05}, namely, the quantization of magnetic
flux piercing the superconducting circuit, being the multiplicity of
the flux quantum $\Phi_0=h/(2e)$ \cite{Dea61,Dol61}, and the {\em ac}
Josephson effect, with the universal frequency-to-voltage ratio given by
$2e/h=1/\Phi_0$ \cite{Jos74}.
Later, with the advent of semiconducting heterostructures \cite{Imr02},
came the quantum Hall effect \cite{And75,Kli80,Lau81,Tsu82,Nov05,Zha05}
and the conductance quantization \cite{Wee88}, bringing us with the
conductance quantum $g_0=se^2/h$ (with the degeneracy $s=1$, $2$, or $4$).
Further development of nanosystems led to the observation of Aharonov-Bohm
effect manifesting itself by magnetoconductance oscillations with the
period $2\Phi_0=h/e$ \cite{Web85}, as well as the universal conductance
fluctuations \cite{Lee85,Alt86,Pal12,YHu17}, characterized
by a~variance $\propto\beta^{-1}(se^2/h)^2$, with an additional
symmetry-dependent prefactor ($\beta=1$, $2$, or $4$). 
Related, but slightly different issue concerns the Wiedemann-Franz (WF) law 
defining the Lorentz number, ${\cal L}_0=\frac{\pi^2}{3}(k_B/e)^2$
(with the Boltzmann constant $k_B$) \cite{Kit05},  
as the proportionality coefficient between electronic part of the
thermal conductivity and electrical conductivity multiplied by absolute
temperature.
Although the WF law is followed, with a~few-percent accuracy, in various
condensed-matter systems, it has never been shown to have metrological
accuracy \cite{Sha03,Mah13,Yos15,Cro16,Ryc21a,Tin23}. 

Some new 'magic numbers' similar to the mentioned above have arrived with
the discovery of graphene, an atomically-thin form of carbon
\cite{Nov05,Zha05}. For undoped graphene samples, charge transport
is dominated by transport via evanescent modes \cite{Kat20}, resulting in
the universal {\em dc} conductivity $4e^2/(\pi{}h)$ accompanied
by the sub-Poissonian shot noise, with a~Fano factor $F=1/3$
\cite{Kat06,Two06,Mia07,Son08,Dan08,Lai16}.
For high frequencies, {\em ac} conductivity is given by $\pi{}e^2/(2h)$, 
leading to the quantized visible light opacity $\pi\alpha$ (with
$\alpha\approx{}1/137.036$ being the fine-structure constant)
\cite{Nai08,Sku10,Mer16}.
A~possible new universal value is predicted for the maximum absolute
thermopower, which approaches $\approx{}k_B/e$ near the charge neutrality
point, for both monolayer and gapless bilayer graphene 
\cite{Wan11,Sus18,Sus19,Zon20,Jay21}.  

\begin{figure*}[!t]
  \includegraphics[width=0.9\linewidth]{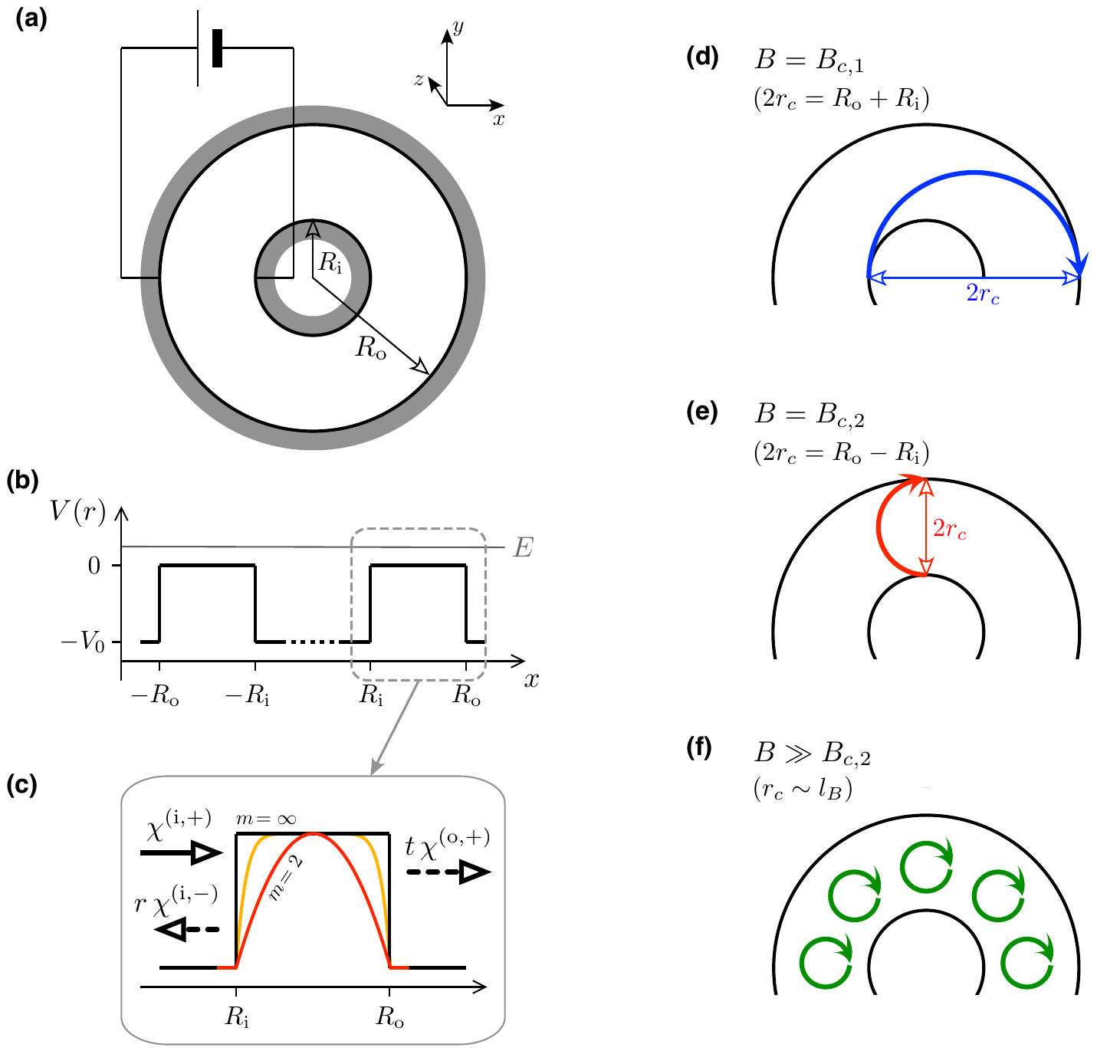}
  \caption{ \label{setup6pan:fig}
    (a) Schematic of Corbino disk in graphene, with the inner radius
    $R_{\rm i}$ and the outer radius $R_{\rm o}$, contacted by two circular
    electrodes (dark areas). A voltage source drives a current through
    the disk. A separate gate electrode (not shown) allows us to tune
    the carrier concentration around the neutrality point. The coordinate
    system $(x,y,z)$ is also shown. 
    (b) Cross section of the electrostatic potential profile given by
    Eq.\ (\ref{v0mpot}) with $m\rightarrow\infty$ (i.e., the rectangular
    barrier) at $y=z=0$.
    (c) Zoom-in of a~single barrier, for $x>0$, showing also the profiles
    for $m=2$ and $8$, with symbolic representations of the incident and
    reflected waves in inner electrode ($x<R_{\rm i}$) and the transmitted wave
    in outer electrode ($x>R_{\rm o}$) with the amplitudes $r$ and $t$
    corresponding to the Fermi energy $E>0$. 
    (d)--(f) Characteristic values of the magnetic field ${\bf B}=(0,0,B)$
    separating different transport regimes.
    At $B=B_{c,1}$, the cyclotron diameter $2r_c=R_{\rm o}+R_{\rm i}$, and
    the particle leaving the inner lead approaches the outer lead regardless
    the initial direction (d).
    At $B=B_{c,2}$, we have $2r_c=R_{\rm o}-R_{\rm i}$, and only the trajectory
    tangent to the inner lead reaches the outer lead (e).
    For higher fields, classical trajectories do not contribute to the
    charge transport, which is possibly only if the resonance with Landau
    level occurs for $E\approx{}E_{n{\rm LL}}$, with $n=0,\pm{}1,\pm{}2,\dots$
    (f). 
  }
\end{figure*}

Away from the charge-neutrality point, ballistic graphene samples show
the sub-Sharvin charge transport \cite{Ryc21b,Ryc22}, characterized by the
conductance reduced by a factor of $\pi/4$ compared to standard Sharvin
contacts in two-dimensional electron gas (2DEG) \cite{Sha65,Bee91}.
What is more, the shot noise is enhanced (comparing to 2DEG) up to
$F\approx{}1/8$ far from the charge-neutrality \cite{Dan08,Lai16}.
Detailed dependence of the above-mentioned factors on a~sample geometry was
recently discussed in analytical terms \cite{Ryc22}, on the example of
edge-free (Corbino) setup, characterized by the inner radius $R_{\rm i}$
and the outer radius $R_{\rm o}$ (see Fig.\ \ref{setup6pan:fig}).
It is further found in Refs.\ \cite{Ryc21b,Ryc22} that the ballistic values
of the conductance and Fano factor are gradually restored when the
potential barrier, defining a~sample area in the effective
Dirac-Weyl Hamiltonian, 
evolves from a~rectangular toward a~parabolic shape. 

Here, we focus on the Corbino geometry, which is often considered
when discussing fundamental aspects of graphene
\cite{Kat20,Ryc09,Ryc10,Pet14,Abd17,Zen19,Sus20,Kam21,Ryc20,Bou22,Kum22,Ryc23}.
In this geometry, charge transport at high magnetic fields is unaffected
by edge states, allowing one to probe the bulk transport properties
\cite{Zen19,Sus20,Kam21,Ryc20,Bou22,Kum22}. Recently, we have shown numerically
that thermoelectric properties in such a~situation are determined by
the energy interval separating consecutive Landau levels rather then by the
transport gap (being the energy interval, for which the cyclotron diameter
$2r_c<R_{\rm o}-R_{\rm i}$) \cite{Ryc23}. 
In this paper, we address a~question how the shot-noise behaves when
the tunneling conductance regime is entered by increasing magnetic field at
a~fixed doping (or decreasing the doping at a~fixed field)?
Going beyond the linear-response regime, we find that the threshold
voltage $U_{\rm on}$, defined as a~source-drain voltage difference that
activates the current at minimal doping, is accompanied by quasi-universal 
(i.e., weakly-dependent on the radii ratio $R_{\rm o}/R_{\rm i}$) value of
$F\approx{}0.56$.
The robustness of the effect is also analyzed when smoothing the electrostatic
potential barrier.

The paper is organized as follows. In Sec.\ \ref{modmet} we briefly present
the effective Dirac Hamiltonian and the numerical approach applied
in remaining parts of the paper. 
In Sec.\ \ref{appcondfa}, we derive an approximation for the transmission
through a~doped Corbino disk at non-zero magnetic field and subsequent
formulas for charge-transfer characteristics: the conductance and
the Fano factor.
Our numerical results, for both the rectangular and smooth potential
barriers, are presented in Sec.\ \ref{resdis}. 
The conclusions are given in Sec.\ \ref{conclu}.

\section{Model and methods}
\label{modmet}

\subsection{Dirac equation for the disk geometry}
Our analysis of the device shown schematically in Fig.\ \ref{setup6pan:fig}
starts from effective wave equation for Dirac fermions  in graphene,
near the $K$ valley, 
\begin{equation}
  \label{direqvr}
  \left[
    v_F\,({\boldsymbol p}+e{\boldsymbol A})\cdot
    {\boldsymbol\sigma} + V(r)
  \right]
  \Psi=E\Psi, 
\end{equation}
where the Fermi velocity is given by $v_F=\sqrt{3}\,t_0a/(2\hbar)$,
with $t_0=2.7\,$eV the nearest-neighbor hopping integral and $a=0.246\,$nm
the lattice parameter, ${\boldsymbol p}=-i\hbar\,(\partial_x,\partial_y)$ is
the in-plane momentum operator, we choose the symmetric gauge
$\mbox{\boldmath$A$}=\frac{B}{2}(-y,x)$ corresponding to the perpendicular,
uniform magnetic field ${\boldsymbol B}=(0,0,B)$, and
$\mbox{\boldmath$\sigma$}=(\sigma_x,\sigma_y)$, where $\sigma_j$ are the
Pauli matrices \cite{vfhbfoo}. The electrostatic potential energy in Eq.\
(\ref{direqvr}), $V(r)$, is given by
\begin{widetext}
\begin{equation}
  \label{v0mpot}
  V(r) = -V_0 \times
  \begin{cases}
    \,\frac{2^m|r-R_{\rm av}|^m}{|R_{\rm o}-R_{\rm i}|^m}
    &  \text{if }\ |r\!-\!R_{\rm av}| \leqslant \frac{R_{\rm o}-R_{\rm i}}{2}, \\
    \,1  &  \text{if }\ |r\!-\!R_{\rm av}| > \frac{R_{\rm o}-R_{\rm i}}{2}, 
  \end{cases}
\end{equation}
\end{widetext}
where we have defined $R_{\rm av}=(R_{\rm i}+R_{\rm o})/2$.
In particular, the limit of $m\rightarrow{}\infty$ corresponds to the
rectangular barrier (with a~cylindrical symmetry); any finite $m\geqslant{}2$
defines
a~smooth potential barrier, interpolating between the parabolic ($m=2$)
and rectangular shape. In principle, barrier smoothing can be regarded 
as a~feature of a~self consistent solution originating from the diffusion
of carriers; we expect this feature to strongly depend on the experimental
details, with graphene-on-hBN devices \cite{Zen19} showing rectangular,
rather then smooth, profiles. 

Symmetry of the problem allows one to look for the wave function in the form
\begin{equation}
  \Psi_j(r,\varphi)=e^{i(j-1/2)\varphi}
  \begin{pmatrix} \chi_a \\ \chi_be^{i\varphi} \end{pmatrix}, 
\end{equation}
where $j=\pm{1/2},\pm{3/2},\dots$ is the total angular-momentum
quantum number, the components $\chi_a=\chi_a(r)$, $\chi_b=\chi_b(r)$,
and we have introduced the polar coordinates $(r,\varphi)$.
Substituting the above into Eq.\ (\ref{direqvr}) bring us to the  system
of ordinary differential equations
\begin{align}
  \chi_a' &= \left(\frac{j-1/2}{r}+\frac{eBr}{2\hbar}\right)\chi_a
  +i\,\frac{E-V(r)}{\hbar{}v_F}\chi_b,
  \label{phapri} \\
  \chi_b' &= i\,\frac{E-V(r)}{\hbar{}v_F}\chi_a
  -\left(\frac{j+1/2}{r}+\frac{eBr}{2\hbar}\right)\chi_b,
  \label{phbpri} 
\end{align}
where primes denote derivatives with respect to $r$.

\subsection{Analytic solutions}
For the disk area, $R_{\rm i}<r<R_{\rm o}$, Eqs.\ (\ref{phapri}), (\ref{phbpri})
typically need to be integrated numerically; key details of the procedure 
are presented in Appendix~\ref{appnumoma}. Here we focus on
the special case of rectangular barrier ($m=\infty$), for which some
analytic solutions were reported \cite{Ryc09,Ryc10,Rec09}.

In particular, in the absence of
magnetic field ($B=0$), the spinors $\chi_j=(\chi_a,\chi_b)^T$ corresponding
to different $j$-s can be written as linear combinations \cite{Ryc09}
\begin{equation}
  \label{chdiskbzer}
  \chi_j^{({\rm disk})}=A_j
  \begin{pmatrix}H_{j-1/2}^{(2)}(kr)\\ i\eta{}H_{j+1/2}^{(2)}(kr)\end{pmatrix}
  + B_j
  \begin{pmatrix}H_{j-1/2}^{(1)}(kr)\\ i\eta{}H_{j+1/2}^{(1)}(kr)\end{pmatrix}, 
\end{equation}
where $H_\nu^{(1)}(\rho)$ [$H_\nu^{(2)}(\rho)$] is the Hankel function of the
first [second] kind, $k=|E|/(\hbar{}v_F)$, the doping sign
$\eta=\mbox{sgn}\,E=\pm{}1$ (with $\eta=+1$ indicating electron doping
and $\eta=-1$ indicating hole doping),
and $A_j$, $B_j$, are arbitrary complex coefficients. 
For $B>0$, Eq.\ (\ref{chdiskbzer}) is replaced by \cite{Ryc10,Rec09}
\begin{equation}
  \label{chdiskbnon}
  \chi_j^{(\mathrm{disk})}=A_j\left(\begin{array}{c}\xi_{j\uparrow}^{(1)} \\ 
     i\eta{}z_{j,1}\xi_{j\downarrow}^{(1)}\end{array}\right)+
  B_j\left(\begin{array}{c} \xi_{j\uparrow}^{(2)} \\ 
    i\eta{}z_{j,2}\xi_{j\downarrow}^{(2)}\end{array}\right),
\end{equation}
where $z_{j,1}=[2(j+s_j)]^{-2s_j}$, $z_{j,2}=2(\beta/k^2)^{s_j+1/2}$ (with $s_j\equiv\frac{1}{2}\mbox{sgn}\,j$, $\beta=eB/(2\hbar)$), and
\begin{equation}
\label{xisnu}
  \xi_{js}^{(\nu)}=e^{-\beta{r}^2/2}(kr)^{|l_s|}\left\{\begin{array}{cc}
    M(\alpha_{js},\gamma_{js},\beta{r}^2), & \nu\!=\!1, \\ 
    U(\alpha_{js},\gamma_{js},\beta{r}^2), & \nu\!=\!2, \\
  \end{array}\right.
\end{equation}
with $l_s=j\mp\frac{1}{2}$ for $s=\uparrow,\downarrow$, $\alpha_{js}=\frac{1}{4}[2(l_{-s}+|l_s|+1)-k^2/\beta]$, and $\gamma_{js}=|l_s|+1$. $M(a,b,z)$ and $U(a,b,z)$ are the confluent hypergeometric functions \cite{Abram}. 

For the leads,  $r<R_{\rm i}$ or $r>R_{\rm o}$, the electrostatic
potential energy is constant, $V(r)=-V_0$. We further assume $B=0$ and
$E>-V_0$ (electron doping) in the leads, allowing one to adapt the wave
function given by Eq.\ (\ref{chdiskbzer}); i.e., for the inner lead, 
$r<R_{\rm i}$,
\begin{equation}
  \label{chinner}
  \chi_j^{({\rm inner})}=
  \begin{pmatrix}H_{j-1/2}^{(2)}(Kr)\\ iH_{j+1/2}^{(2)}(Kr)\end{pmatrix}
  + r_j
  \begin{pmatrix}H_{j-1/2}^{(1)}(Kr)\\ iH_{j+1/2}^{(1)}(Kr)\end{pmatrix}, 
\end{equation}
and for the outer lead, $r>R_{\rm o}$, 
\begin{equation}
  \label{chouter}
  \chi_j^{({\rm outer})}= t_j
  \begin{pmatrix}H_{j-1/2}^{(2)}(Kr)\\ iH_{j+1/2}^{(2)}(Kr)\end{pmatrix}, 
\end{equation}
where $K=|E+V_0|/(\hbar{}v_F)$ and we have introduced the reflection and
transmission coefficient.
The first spinor in each of Eqs.\ (\ref{chinner}) and (\ref{chouter})
represents the incoming (i.e., propagating from $r=0$) wave, the
second spinor in Eq.\ (\ref{chinner}) represents the outgoing
(propagating from $r=\infty$) wave.

\subsection{Mode-matching method}

Since the current-density operator following from Eq.\ (\ref{direqvr}),
${\boldsymbol j}=ev_F{\boldsymbol\sigma}$ does not involve differentiation,
the mode-matching conditions for $r=R_{\rm i}$ and $r=R_{\rm o}$ reduce to
the equalities for spinor components, namely
\begin{equation}
  \chi_j^{({\rm inner})}(R_{\rm o})=\chi_j^{({\rm disk})}(R_{\rm o})
  \ \ \text{and }\ 
  \chi_j^{({\rm disk})}(R_{\rm i})=\chi_j^{({\rm outer})}(R_{\rm i}). 
\end{equation}
The resulting formula for transmission probability for $j$-th mode becomes
particularly simple upon taking the limit of heavily doped leads,
$U_0\rightarrow{}\infty$. In particular, for $B=0$, substituting
Eq.\ (\ref{chdiskbzer}) into the above gives \cite{wronfoo}
\begin{equation}
\label{tjhank}
  T_j=|t_j|^2=
  \frac{16}{\pi^2{}k^2{}R_{\rm i}{}R_{\rm o}}\,
  \frac{1}{\left[\mathfrak{D}_{j}^{(+)}\right]^2
    + \left[\mathfrak{D}_{j}^{(-)}\right]^2},
\end{equation}
where
\begin{align}
\label{ddnupm}
  \mathfrak{D}_{j}^{(\pm)} &= \mbox{Im}\left[ 
    H_{j-1/2}^{(1)}(kR_{\rm i})H_{j\mp{}1/2}^{(2)}(kR_{\rm o})\right. \nonumber\\
    & \ \ \ \ \ \ \ \ \ \ 
    \pm \left.H_{j+1/2}^{(1)}(kR_{\rm i})H_{j\pm{}1/2}^{(2)}(kR_{\rm o})
    \right]. 
\end{align}

Analogously, for $B>0$ one finds, using Eqs.\ (\ref{chdiskbnon}) and
(\ref{xisnu}),  
\begin{equation}
\label{tjbnon}
  T_j=|t_j|^2=
  \frac{16\,(k^2/\beta)^{|2j-1|}}{k^2R_\mathrm{i}R_\mathrm{o}%
  \,(X_j^2+Y_j^2)}
  \left[\frac{\Gamma(\gamma_{j\uparrow})}{\Gamma(\alpha_{j\uparrow})}\right]^2,
\end{equation}
where $\Gamma(z)$ is the Euler Gamma function, and
\begin{align}
& X_j = w_{j\uparrow\uparrow}^- + z_{j,1}z_{j,2}w_{j\downarrow\downarrow}^-,\ \ \ 
  Y_j = z_{j,2}w_{j\uparrow\downarrow}^+ - z_{j,1}w_{j\downarrow\uparrow}^+, \nonumber\\
& w_{jss'}^\pm= \xi_{js}^{(1)}(R_\mathrm{i})\xi_{js'}^{(2)}(R_\mathrm{o})
  \pm \xi_{js}^{(1)}(R_\mathrm{o})\xi_{js'}^{(2)}(R_\mathrm{i}). 
\end{align}
For $B<0$, one gets $T_j(B)=T_{-j}(-B)$. 

Details of numerical mode-matching, applicable for smooth potentials, 
are given in Appendix~\ref{appnumoma}.

\subsection{Landauer-B\"{u}ttiker formalism}

In case the nanoscopic system is connected to external reservoirs,
characterized by the electrochemical potentials $\mu$ and $\mu+eU_{\rm eff}$
(for simplicity, the two reservoirs are considered; for more general
discussion see Ref.\ \cite{Naz09}), the conductance of
the system is related to the transmission probabilities for normal modes
($T_j$-s) via 
\begin{equation}
\label{guefflan}
  G(U_{\rm eff}) = \frac{\langle{}I\rangle}{U_{\rm eff}}=
  \frac{g_0}{U_{\rm eff}}
  \int_{\mu}^{\mu+eU_{\rm eff}}d\epsilon\sum_jT_j(\epsilon), 
\end{equation}
where $\langle{}I\rangle$ denotes the average electric current and the
zero-temperature limit is taken.
The conductance quantum is $g_0=4e^2/h$, taking into account spin and valley
degeneracies.
$U_{\rm eff}$ denotes the effective voltage difference between the reservoirs
(notice that the actual voltage applied may differ from $U_{\rm eff}$ due to
charge-screening effects).
Similarly, the Fano factor, relating the current variance,
$\left\langle{}\left(I-\langle{}I\rangle\right)^2\right\rangle$,
to the value
$\left\langle{}\left(I-\langle{}I\rangle\right)^2\right\rangle_{\rm Poisson}$
one would measure in the absence of correlations between scattering events
(occurring, e.g., in the tunneling limit of $T_j\ll{}1$ for all $j$-s), 
is given by
\begin{widetext}
\begin{equation}
\label{fuefflan}
  F(U_{\rm eff}) =
  \frac{\left\langle{}\left(I-\langle{}I\rangle\right)^2\right\rangle}{\left\langle{}\left(I-\langle{}I\rangle\right)^2\right\rangle_{\rm Poisson}} =
  \frac{g_0}{GU_{\rm eff}}
  \int_{\mu}^{\mu+eU_{\rm eff}}d\epsilon\sum_j
  T_j(\epsilon)\left[1-T_j(\epsilon)\right], 
\end{equation}
\end{widetext}
where
$\left\langle{}\left(I-\langle{}I\rangle\right)^2\right\rangle_{\rm Poisson}
=e\langle{}I\rangle/\Delta{}t=eGU_{\rm eff}/\Delta{}t$
with $\Delta{}t$ denoting the time of measurement. 

In the linear-response regime ($U_{\rm eff}\rightarrow{}0$), Eqs.\ 
(\ref{guefflan}) and (\ref{fuefflan}) reduces to
\begin{equation}
\label{gueff0}
  G(U_{\rm eff}\rightarrow{}0) = g_0\sum_jT_j,
\end{equation}
and
\begin{equation}
\label{fueff0}
  F(U_{\rm eff}\rightarrow{}0) = \frac{\sum_jT_j(1-T_j)}{\sum_jT_j}, 
\end{equation}
where $T_j=T_j(\mu)$. For the disk geometry, summation range is limited
by the number of propagating modes in the inner lead,
$|j|\leqslant{}j_{\rm max}=\lfloor{}KR_{\rm i}\rfloor-\frac{1}{2}$
with $\lfloor{}x\rfloor$ denoting the floor function of $x$. 
(For heavily-doped leads, $j_{\rm max}\rightarrow{}\infty$.)

As a~notable example, we consider the zero-doping limit
($\mu\rightarrow{}0$). 
In such a~case, Eq.\ (\ref{tjbnon}) simplifies to \cite{Ryc10,Kat10}
\begin{equation}
  T_j(\mu\rightarrow{}0)=
  \frac{1}{\cosh^2[(j+\Phi/\Phi_0)\ln(R_{\rm o}/R_{\rm i})]}, 
\end{equation}
where $\Phi=\pi(R_\mathrm{o}^2-R_\mathrm{i}^2)B$ is the flux piercing the disk
area and we have defined $\Phi_0=2\,(h/e)\ln(R_{\rm o}/R_{\rm i})$.
Assuming the narrow-disk range, $R_{\rm o}\approx{}R_{\rm i}$, we can
approximate the sums occurring in Eqs.\ (\ref{gueff0}) and (\ref{fueff0})
by integrals, obtaining
\begin{equation}
  G\approx{}G_{\rm diff}=\frac{2\pi{}\sigma_0}{\ln(R_{\rm o}/R_{\rm i})}
  \ \ \ \
  \text{and }
  \ \ \ \
  F\approx{}F_{\rm diff}=\frac{1}{3}. 
\end{equation}
The above reproduces pseudodiffusive conductance and the shot-noise power
for a~disk geometry \cite{Ryc09}. For larger $R_{\rm o}/R_{\rm i}$, both
characteristics are predicted to show 
approximately sinusoidal conductance oscillations with the field $B$
\cite{Ryc10,Kat10,Ryc12}.

The case of doped disk, for which one may expect to observe some features
of the sub-Sharvin charge transport \cite{Ryc21b,Ryc22}, is discussed next.

\begin{figure*}[!t]
  \includegraphics[width=0.9\linewidth]{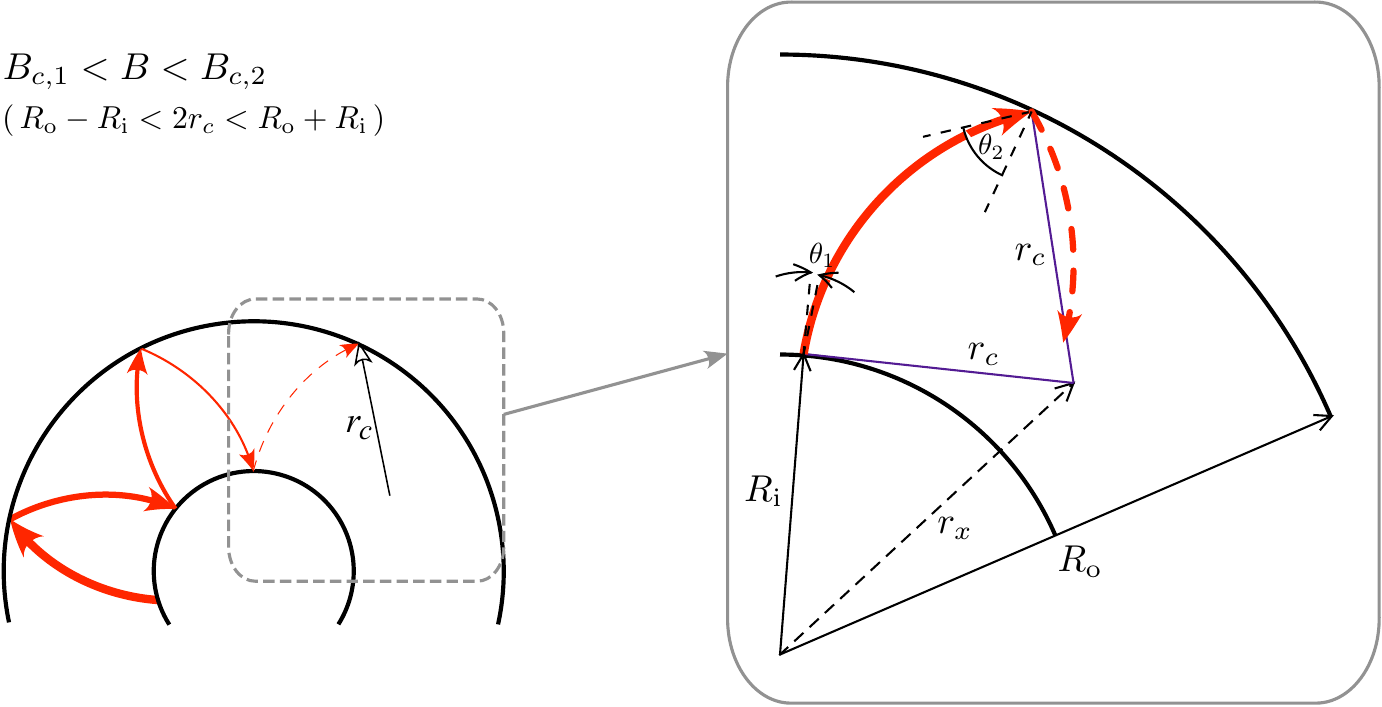}
  \caption{ \label{scattering2:fig}
    Propagation between consecutive scatterings on interfaces at
    $r=R_{\rm i}$ and $r=R_{\rm o}$ in a uniform magnetic field 
    $B_{c,1}<B<B_{c,2}$. A zoom-in shown an arc of single cyclotron orbit
    centered at $r = r_x$, with its radii $r_c$, and incident angles
    $\theta_1$ (for $r=R_{\rm i}$) and $\theta_2$ (for $r=R_{\rm o}$). 
  }
\end{figure*}

\section{Approximate conductance and Fano factor at the magnetic field}
\label{appcondfa}

Before calculating the conductance $G$ and Fano factor $F$ within the
mode-matching method described in Sec.\ \ref{modmet}, we first present
the approximating formulas for incoherent transport, obtained by
adapting the derivation of Ref.\ \cite{Ryc22} for the $B>0$ case.

\subsection{Corbino disk in graphene as a~double barrier}
A key step in the derivation is to observe that, in the multimode regime
($kR_{\rm i}\gg{}1$) for which one can consider well-defined trajectories, 
the disk symmetry cause that incident angles $\theta_1$ and $\theta_2$,
corresponding to the interfaces at $r=R_{\rm i}$ and $r=R_{\rm o}$
(see Fig.\ \ref{scattering2:fig}) remain constant (up to a sign) after
multiple scatterings. Therefore, one can apply the double-contact formula
for incoherent transmission
\cite{Dat97,intfoo}, namely
\begin{widetext}
\begin{align}
  \left\{T\right\}_{\rm incoh} &= \frac{1}{2\pi}\int_{-\pi}^{\pi}d\phi\, 
  \frac{T_1T_2}{2-T_1-T_2+T_1{}T_2-2\sqrt{(1-T_1{})(1-T_2)}\cos{\phi}} =
  \frac{T_1T_2}{T_1+T_2-T_1T_2},
  \label{tticoh}
\end{align}
\end{widetext}
where the transmission probabilities $T_1$, $T_2$, corresponding to a~potential
step of infinite height, are given by
\begin{equation}
\label{ttlthl}
  T_l = \frac{2\cos\theta_l}{1+\cos\theta_l}, \ \ \ \ \ \ l=1,2, 
\end{equation}
and $\phi$ is assumed to be a~random phase acquired during the propagation
between $r=R_{\rm i}$ and $r=R_{\rm o}$ (or vice versa).
Similarly, we calculate the incoherent squared transmission, useful
when evaluating the Fano factor,
\begin{widetext}
\begin{align}
  \left\{T^2\right\}_{\rm incoh} &= \frac{1}{2\pi}\int_{-\pi}^{\pi}d\phi\,
  \left(\frac{T_1T_2}{2-T_1-T_2+T_1{}T_2-2\sqrt{(1-T_1{})(1-T_2)}
  \cos{\phi}}\right)^2
  = \frac{(T_1T_2)^2(2-T_1-T_2+T_1T_2)}{(1+T_1T_2-T_1T_2)^3}. 
  \label{tt2icoh}
\end{align}
\end{widetext}

Next, the incoherent conductance in the linear-response regime is evaluated
by inserting $\left\{T\right\}_{\rm incoh}$ (\ref{tticoh}) into
Eq.\ (\ref{gueff0}),
\begin{equation}
\label{gincoh}
  G_{\rm incoh} = G_{\rm Sharvin}
  \left\langle{}\left\{T\right\}_{\rm incoh}\right\rangle_{u=\sin\theta_1}, 
\end{equation}
with
\begin{equation}
  G_{\rm Sharvin}=2g_0kR_{\rm i}.  
\end{equation}
For the Fano factor, one can analogously derive from Eq.\ (\ref{fueff0}) 
\begin{equation}
\label{fincoh}
  F_{\rm incoh}=1-\frac{\left\langle{}\left\{T^2\right\}_{\rm incoh}\right\rangle_{u=\sin\theta_1}}{\left\langle{}\left\{T\right\}_{\rm incoh}\right\rangle_{u=\sin\theta_1}}. 
\end{equation}

The summation over $2kR_{\rm i}$ modes is approximated in Eqs.\ (\ref{gincoh}), 
(\ref{fincoh})
by averaging over the variable $u=\sin\theta_1$, within the range of
$-1\leqslant{}u\leqslant{}1$. 
Explicitly,
\begin{equation}
\label{ttnicohav}
  \left\langle{}\left\{T^n\right\}_{\rm incoh}\right\rangle_{u=\sin\theta_1} = 
  \frac{1}{2}\int_{u_c}^{1}du\left\{T^n\right\}_{\rm incoh},
  \ \ \ \ n=1,2,
\end{equation}
where the lower integration limit ($u_c$) is defined via the value of
$\sin\theta_1$, below which the trajectory cannot reach the outer interface
($r=R_{\rm o}$). (In other words, for $u=\sin\theta_1<u_c$, the geometric
derivation to be presented below leads to $|\sin\theta_2|>1$.)

The missing elements, necessary to calculate
$\left\langle{}\left\{T^n\right\}_{\rm incoh}\right\rangle_{u=\sin\theta_1}$
in Eq.\ (\ref{ttnicohav}) is the dependence of $\theta_2$ on $\theta_1$
and $B$ [see Eqs.\ (\ref{tticoh}), (\ref{ttlthl}), and (\ref{tt2icoh})],
as well as the dependence of $u_c$ on $B$.
Since we have assumed constant electrostatic potential energy in the disk
area, the trajectory between subsequent scatterings
(see Fig.\ \ref{scattering2:fig}) forms an arc, with the constant radii
\begin{equation}
\label{rcdef}
r_c=\hbar{}k/(eB)=|E|/(v_FeB), 
\end{equation}
(i.e., the cyclotron radius for massless Dirac particle at $B>0$),
centered at the distance $r_x$ from the origin.
Now, solving the two triangles with a~common edge $r_x$ (dashed line)
and the opposite vertices in two scattering points, we find
\begin{equation}
\label{rx2tri1}
  r_x^2 = R_{\rm i}^2+r_c^2 + 2R_{\rm i}{}r_c\sin\theta_1 
\end{equation}
(for the triangle containing a~scattering point at $r=R_{\rm i}$), and
\begin{equation}
\label{rx2tri2}
  r_x^2 = R_{\rm o}^2+r_c^2 - 2R_{\rm o}{}r_c\sin\theta_2 
\end{equation}
(for the triangle containing a~scattering point at $r=R_{\rm o}$).
Together, Eqs.\ (\ref{rx2tri1}) and (\ref{rx2tri2}) lead to
\begin{equation}
\label{th2vsu}
  \sin\theta_2 
  = \frac{R_{\rm o}^2-R_{\rm i}^2-2R_{\rm i}r_cu}{2R_{\rm o}r_c}. 
\end{equation}
Subsequently, the value of $u_c$ in Eq.\ (\ref{ttnicohav}) is given by
\begin{equation}
  \label{ucases}
  u_c = \begin{cases}
    -1,  & \text{if }\ B\leqslant{}B_{c,1} \\
     \frac{R_{\rm o}^2-R_{\rm i}^2}{2R_{\rm i}r_c}-\frac{R_{\rm o}}{R_{\rm i}},
     & \text{if }\
     B_{c,1}<B\leqslant{}B_{c,2} \\
    1,  & \text{if }\ B>B_{c,2} \\
  \end{cases},  
\end{equation}
where we have additionally defined
\begin{equation}
  B_{c,m}=\frac{2\hbar{}k}{e\left[R_{\rm o}-(-1)^m{}R_{\rm i}\right]},
  \ \ \ \ \ m=1,2. 
\end{equation}

\subsection{The zero-field limit}
Typically, averages occurring in Eqs.\ (\ref{gincoh}) and (\ref{fincoh})
need to be evaluated numerically. Analytic expressions are available, e.g.,
for zero magnetic field \cite{Ryc22}
\begin{widetext}
\begin{align}
  G_{\rm incoh}(B\!\rightarrow{}\!0) &= G_{\rm Sharvin}\,  
  \frac{(2a+\frac{1}{a})\arcsin{}a+3\sqrt{1-a^2}-\frac{\pi}{2}(a^2+2)}{1-a^2},
  \label{ggzeroicoh} \\
  F_{\rm incoh}(B\!\rightarrow{}\!0) &= 
  \frac{2a\sqrt{1-a^2}(53+279a^2+88a^4)-3\pi{}a(12+82a^2+45a^4+a^6)+6(1+45a^2+82a^4+12a^6)\arcsin{}a}{6(1-a^2)^2\left[\pi{}a(a^2+2)-6a\sqrt{1-a^2}-2(2a^2+1)
  \arcsin{}a\right]}, 
  \label{ffzeroicoh}
\end{align}
\end{widetext}
where we have defined the inverse radii ratio $a=R_{\rm i}/R_{\rm o}$.

\subsection{The zero-conductance limit}
In the this paper, we focus on the limit of $B\rightarrow{}B_{c,2}-$
(i.e., $B$ approaching $B_{c,2}$ from below), for which
$G_{\rm incoh}\rightarrow{}0$. Introducing the dimensionless
$0<\varepsilon\ll{}1$, one can express the cyclotron diameter, 
see Eq.\ (\ref{rcdef}), as
\begin{equation}
  \label{epsdef}
  2r_c = R_{\rm o}-R_{\rm i} + \varepsilon{}\left(R_{\rm o}-R_{\rm i}\right).
\end{equation}
In turn, the value $u_c$, see Eq.\ (\ref{ucases}), can be approximated
(up to the leading order in $\varepsilon$) as
\begin{equation}
  u_c\approx{}1-\varepsilon\left(1+\frac{R_{\rm o}}{R_{\rm i}}\right). 
\end{equation}
It is now convenient to define the variable
\begin{equation}
  \alpha = \frac{1-u}{\varepsilon\left(1+{R_{\rm o}}/{R_{\rm i}}\right)},
\end{equation}
such that the integration over $u_c\leqslant{}u\leqslant{}1$, occurring when
evaluating 
$\left\langle{}\left\{T^n\right\}_{\rm incoh}\right\rangle_{u=\sin\theta_1}$
from Eq.\ (\ref{ttnicohav}), 
can be replaced by integration over $1\geqslant{}\alpha\geqslant{}0$.
Transmission probabilities ($T_1$, $T_2$) for the interfaces at $r=R_{\rm i}$
and $r=R_{\rm o}$, see Eqs.\ (\ref{tticoh}), (\ref{ttlthl}), (\ref{tt2icoh})
and (\ref{th2vsu}), can now be approximated as
\begin{align}
  T_1 &\approx
  2\sqrt{2\alpha\left(1+{\scriptstyle\frac{1}{a}}\right)}\,\varepsilon^{1/2},  \\
  T_2 &\approx
  2\sqrt{2(1-\alpha)\left(1+{a}\right)}\,\varepsilon^{1/2}, 
\end{align}
where we have used $a=R_{\rm i}/R_{\rm o}$ again.

Using the above expressions, we can now rewrite the averages occurring 
in Eq.\ (\ref{ttnicohav}), up to the leading order in $\varepsilon$ again,
as follows
\begin{widetext}
\begin{align}
  \left\langle{}\left\{T\right\}_{\rm incoh}\right\rangle_{u} &=
  \frac{G_{\rm incoh}}{G_{\rm Sharvin}} 
  \approx
  \frac{2\left(1+\frac{1}{a}\right)}{\sqrt{2}}\,\varepsilon^{3/2}
  \int_0^1{}d\alpha
  \frac{\sqrt{\alpha(1+\frac{1}{a})}\sqrt{(1-\alpha)(1+a)}}{\sqrt{\alpha(1+\frac{1}{a})}+\sqrt{(1-\alpha)(1+a)}},
  \label{tt1icoheps}
  \\
  \left\langle{}\left\{T^2\right\}_{\rm incoh}\right\rangle_{u} &\approx
  \frac{2\left(1+\frac{1}{a}\right)}{\sqrt{2}}\,\varepsilon^{3/2}
  \int_0^1{}d\alpha
  \frac{\alpha(1-\alpha)\left(2+\frac{1}{a}+a\right)}{\left[\sqrt{\alpha(1+\frac{1}{a})}+\sqrt{(1-\alpha)(1+a)}\right]^3}.
  \label{tt2icoheps}
\end{align}
\end{widetext}
Remarkably, both quantities decays as $\propto{}\varepsilon^{3/2}$, but their
ratio, occurring in Eq.\ (\ref{fincoh}) for the Fano factor, remains constant
(for a~given $a$).
The integrals in Eqs.\ (\ref{tt1icoheps}) and (\ref{tt2icoheps}) can be
calculated analytically, leading to 
\begin{widetext}
\begin{align}
  F_{\rm incoh}(B\!\rightarrow{}\!B_{c,2}-) &=
  1 - \left\{
  4\sqrt{a} + 39a - 58a^{3/2} - 23a^2 - 23a^{5/2} - 58a^{3} + 39a^{7/2} + 4a^4
  \right. \nonumber\\
  & \hspace{8ex}  + \left. 
  \left(69a^2 - 18a^3 - 18a\right)\sqrt{1+a}
  \,\mbox{artanh}\left(\frac{\sqrt{a}}{\sqrt{1+a}}\right)
  \right. \nonumber\\
  & \hspace{8ex} + \left. 
  \left(69a^2 - 18a - 18a^3\right)\sqrt{1+a}
  \,\mbox{artanh}\left(\frac{1}{\sqrt{1+a}}\right)
  \right\}
  \nonumber\\
  &{\Big/} \left\{
  (1+a)^3\left(1+\sqrt{a}\right)\left(1-3\sqrt{a}+a\right)
  +3a(1+a)^{5/2}\,\mbox{artanh}\left[\frac{(1+\sqrt{a})\sqrt{1+a}}{1+\sqrt{a}+a}\right]
  \right\}.
  \label{fficohbc2}
\end{align}
\end{widetext}

Numerical values of $F_{\rm incoh}(B\!\rightarrow{}\!0)$ and
$F_{\rm incoh}(B\!\rightarrow{}\!B_{c,2}-)$ for selected $a=R_{\rm i}/R_{\rm o}$ 
are given in Table~\ref{fanoriro}.
For $F_{\rm incoh}(B\!\rightarrow{}\!B_{c,2}-)$, 
we see that the poissonian value of $F_{\rm incoh}=1$, which one could expect
due to the vanishing conductance, is reconstructed only for $a\rightarrow{}0$
(i.e., for $R_{\rm o}\gg{}R_{\rm i}$).
For finite radii ratios, nontrivial values of $0<F_{\rm incoh}<1$ occurs. 
Remarkable, for moderate disk proportions ($a\geqslant{}0.5$),
$F_{\rm incoh}(B\!\rightarrow{}\!B_{c,2}-)$ shows very weak dependence on $a$,
decaying by less then $2\%$ (from $F_{\rm incoh}\approx{}0.56$ at $a=0.5$ to
$F_{\rm incoh}\approx{}0.55$ for $a\rightarrow{}1$, with  $a\rightarrow{}1$
representing the narrow-disk limit of $R_{\rm o}\approx{}R_{\rm i}$). 

For this reason, in the following numerical analysis, we fixed the disk
radii ratio at $a=0.5$ (i.e., $R_{\rm o}=2R_{\rm i}$).
We also stress that the derivation presented above, holds true for the
parameter $\varepsilon\rightarrow{}0+$, quantifying the ratio of
cyclotron diameter $2r_c$ to radii difference $R_{\rm o}-R_{\rm i}$, see Eq.\
(\ref{epsdef}). Therefore, it is irrelevant whether one increases the
magnetic field at a~fixed chemical potential, or reduces the chemical
potential at a~fixed $B>0$ (as long as the system stays in a~multimode range,
$kR_{\rm i}\gg{}1$).

\begin{table}[!hb]
\caption{
  Selected numerical values of $F_{\rm incoh}(B\!\rightarrow{}\!0)$, see
  Eq.\ (\ref{ffzeroicoh}), and $F_{\rm incoh}(B\!\rightarrow{}\!B_{c,2}-)$,
  see Eq.\ (\ref{fficohbc2}).
  Box marks the values for $a=0.5$ (i.e., $R_{\rm o}=2R_{\rm i}$) to be compared
  with the results following from numerical simulations of quantum
  transport presented Sec.\ \ref{resdis}. 
  \label{fanoriro}
}
\begin{tabular}{ccc}
\hline\hline
  $\ a=R_{\rm i}/R_{\rm o}\ $  & $\ F_{\rm incoh}(B\rightarrow{}0)\ $ &
    $\ F_{\rm incoh}(B\rightarrow{}B_{c,2}-)\ $ \\
\hline
  0  &  0.106528  &  1  \\
  0.1  &  0.106705  &  0.630994 \\
  0.2  &  0.107239  &  0.591829 \\
  0.3  &  0.108136  &  0.573885 \\
  0.4  &  0.109409  &  0.563905 \\
  {\smash{\fboxsep=0pt\llap{\rlap{\fbox{\strut\makebox[2.25in]{}}}~}}
  \ignorespaces}
  0.5  &  0.111074  &  0.557898 \\
  0.6  &  0.113151  &  0.554178 \\
  0.7  &  0.115663  &  0.551894 \\
  0.8  &  0.118619  &  0.550565 \\
  0.9  &  0.121963  &  0.549899 \\
  1.0  &  0.125000  &  0.549708 \\
\hline\hline
\end{tabular}
\end{table}

\section{Results and discussion}
\label{resdis}

Main doubt arising when we consider the applicability of Eq.\
(\ref{fficohbc2}) for real quantum systems concerns the possible role
of evanescent waves, totally neglected in our derivation.
Obviously, they should not play an important role when the system is
highly conducting (such as in the zero-field case \cite{Ryc22});
however, since the Fano factor is determined by the ratio of two cumulants,
both vanishing for sufficiently high field, it is not fully clear which
contribution (from propagating or from evanescent modes) would govern
the value of $F$ for $B\rightarrow{}B_{c,2}$? On the other hand,
resonances with Landau levels are not expect to play a~significant role,
as they form very narrow transmission peaks, contributions of which
get immediately smeared out beyond the linear-response regime. 

In the remaining parts of the paper, compare the results of computer
simulation of quantum transport through the disk in graphene, with the
predictions for incoherent scattering presented in Sec.\ \ref{appcondfa},  
in attempt to propose an experimental procedure allowing one to extract the
nontrivial value of $F\approx{}0.55$ from the data plagued with other
contributions.

\begin{figure*}[!t]
  \includegraphics[width=\linewidth]{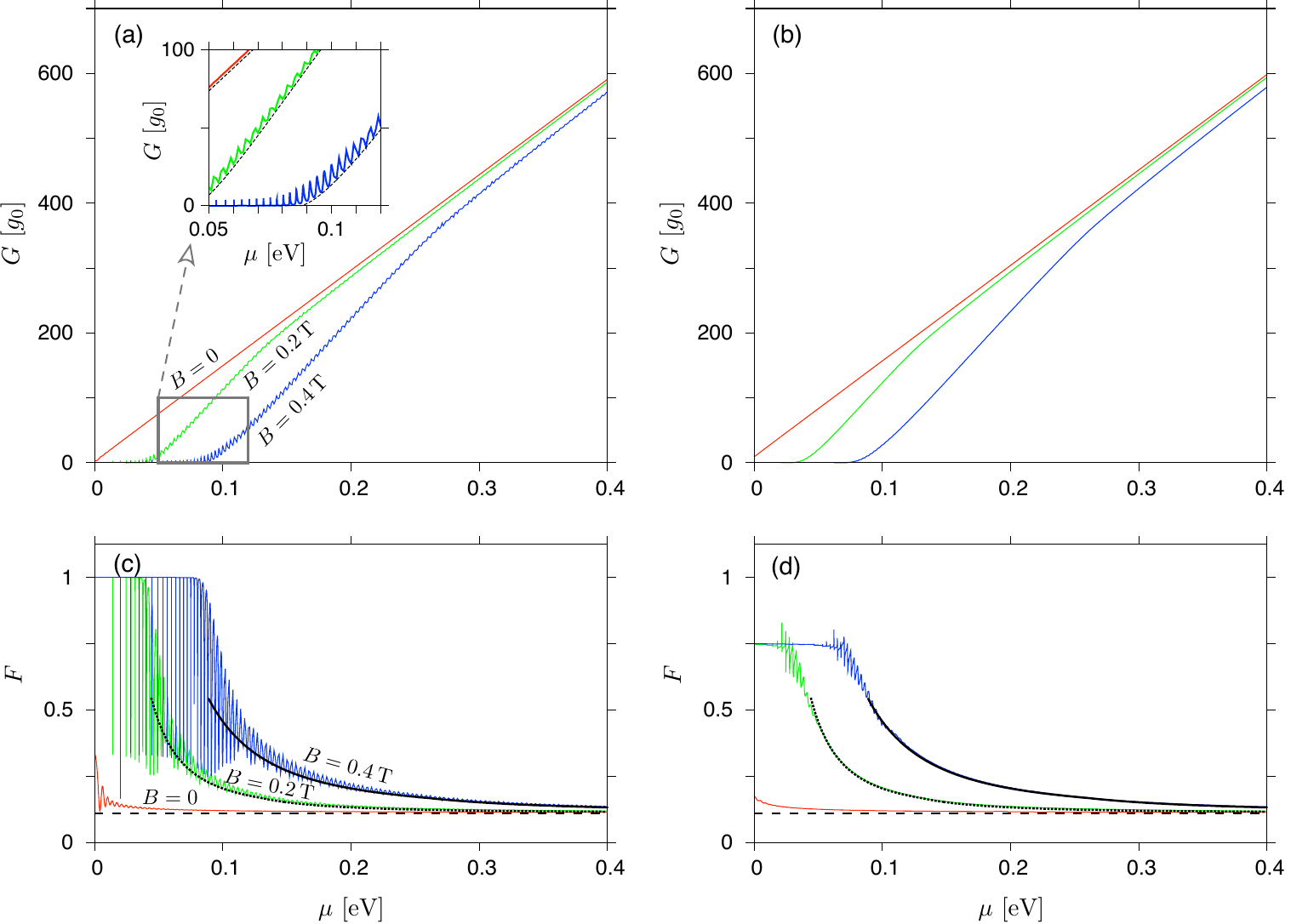}
  \caption{ \label{gfan4pan:fig}
  (a,b) Conductance and (c,d) the Fano factor for the Corbino disk
  in graphene with the radii $R_{\rm o}=2R_{\rm i}=1000\,$nm and the rectangular
  potential barrier (i.e., $V_0\rightarrow{}\infty$ and $m\rightarrow{}\infty$
  in Eq.\ (\ref{v0mpot})) displayed as functions of the chemical potential.
  The values of magnetic field are $B=0$ (red solid lines in all plots),
  $B=0.2\,$T (green solid lines), and $B=0.4\,$T (blue solid lines).
  Inset in (a) is a~zoom-in, with black dashed lines depicting the
  incoherent conductance, see Eq.\ (\ref{gincoh}).
  (a) and (c) show the linear-response results, see Eqs.\ (\ref{gueff0})
  and (\ref{fueff0});
  the datasets in (b) and (d) are obtained from Eqs.\ (\ref{guefflan})
  and (\ref{fuefflan})  with $U_{\rm eff}=0.01\,$V.
  Remaining lines in (c,d) [black solid, black dotted, and black dashed]
  mark the incoherent Fano factor, see Eq.\
  (\ref{fincoh}); the values of magnetic field are specified for lines in (c),
  and are the same in (d).
  (For $B=0$, horizontal lines mark $F_{\rm incoh}(B\rightarrow{}0)=0.111074$
  corresponding to $R_{\rm o}=2R_{\rm i}$, see Table~\ref{fanoriro}.) 
  }
\end{figure*}

\begin{figure*}[!t]
  \includegraphics[width=\linewidth]{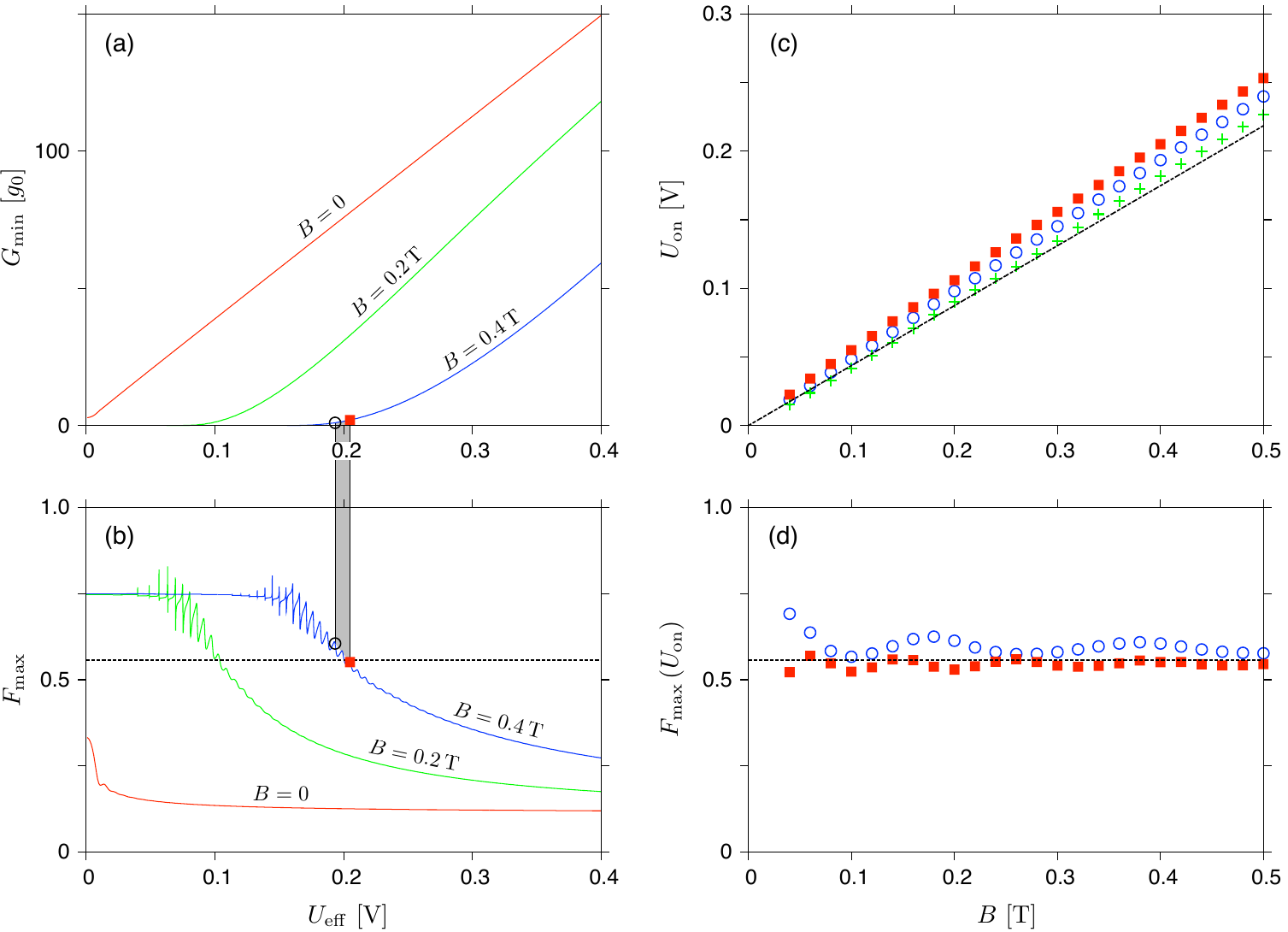}
  \caption{ \label{gmifma4pan:fig}
  (a,b) Minimal conductivity and maximal Fano factor,
  see Eqs.\ (\ref{guefflan}) and (\ref{fuefflan}), corresponding to the
  chemical potential fixed at $\mu=-eU_{\rm eff}/2$, versus the effective
  voltage. The magnetic field is specified for each line.  
  (c) The activation voltage, defined via $G_{\rm min}(U_{\rm on}^{(1)})=g_0$
  [blue open circles], $G_{\rm min}(U_{\rm on}^{(2)})=2g_0$ [red solid squares],
  or obtained from scaling according to Eq.\ (\ref{uon0scal}) 
  [green crosses], displayed versus the magnetic field. 
  (d) The Fano factor corresponding to $U_{\rm eff}=U_{\rm on}$ shown in (c).
  Horizontal dashed lines in (b,d) mark the value of
  $F_{\rm incoh}(B\rightarrow{}B_{c,2}-)=0.557898$ for $R_{\rm o}=2R_{\rm i}$
  (see Table~\ref{fanoriro}). 
  Dashed line in (c) depicts the approximation given in Eq.\ (\ref{uonicoh}). 
  The remaining system parameters are same as in Fig.\ \ref{gfan4pan:fig}.
  }
\end{figure*}

\subsection{The rectangular barrier of an infinite height}
As a~first numerical example, we took the limit of $V_0\rightarrow{}\infty$
and $m\rightarrow{}\infty$ in Eq.\ (\ref{v0mpot}), for which close-form
expressions for transmission probabilities were presented in Sec.\
\ref{modmet}.

In Fig.\ \ref{gfan4pan:fig}, we compare the linear-response conductance
$G(U_{\rm eff}\rightarrow{}0)$, see Eq.\ (\ref{gueff0}), with $G(U_{\rm eff})$
calculated from Eq.\ (\ref{guefflan}) for a~small but nonzero value of
$U_{\rm eff}=0.01\,$V, both displayed as functions of the chemical potential.
Also in Fig.\ \ref{gfan4pan:fig}, same comparison is presented for the
Fano factor $F(U_{\rm eff})$ [see Eqs.\ (\ref{fueff0}) and (\ref{fuefflan})].
It easy to see that prominent, aperiodic oscillations visible for both
charge-transfer cumulants in the $U_{\rm eff}\rightarrow{}0$ limit are
significantly reduced even for small $U_{\rm eff}>0$.
In fact, for $U_{\rm eff}=0.01\,$V and $B>0$, the values of $F_{\rm incoh}$
calculated from Eq.\ (\ref{fincoh}) [black lines] are closely followed
by $F(U_{\rm eff})$ obtained from the numerical mode-matching, as long as
the former can be defined, i.e., for $B<B_{c,2}$ at a~given $\mu$.
We further notice that the value of $\mu$ for which $B\approx{}B_{c,2}$ and
$F(U_{\rm eff})\approx{}0.56$ corresponds to $G(U_{\rm eff})\sim{}g_0$
(up to the order of magnitude). 
For smaller $\mu$, such that $B>B_{c,2}$ and $F_{\rm incoh}$
is undefined, $F(U_{\rm eff})$ saturates near the value $\approx{}0.75$,
apparently below the poissonian limit of $F=1$. 

To better understand the nature of the results we now,
in Fig.\ \ref{gmifma4pan:fig}, go further beyond the linear-response regime,
calculating $G(U_{\rm eff})$ and $F(U_{\rm eff})$ for $\mu=-eU_{\rm eff}/2$
(notice that for an infinite rectangular barrier we have the particle-hole
symmetry, and both cumulants are even upon
$\mu\leftrightarrow{}-\mu+eU_{\rm eff}$) and displaying them as functions
of $U_{\rm eff}$.

We further introduce the activation voltage $U_{\rm on}=U_{\rm on}(B)$,
meaning of which can be understood as follows. 
The cyclotron diameter, see Eq.\ (\ref{rcdef}), naturally defines the range
of energies for which $2r_c(E)<R_{\rm o}-R_{\rm i}$ and the system shows
$G\approx{}0$ (up to the evanescent modes). On the other hand, as we have set
$\mu=-eU_{\rm eff}/2$, the effective voltage defines the energy range of
$|E|\leqslant{}eU_{\rm eff}/2$, being the integration interval in Eqs.\
(\ref{guefflan}) and (\ref{fuefflan}). In turn, $G(U_{\rm eff})>0$ is
expected for $U_{\rm eff}\geqslant{}U_{\rm on}$, a~value of which can be
approximated as
\begin{equation}
  \label{uonicoh}
  U_{{\rm on},{\rm incoh}} = v_F{}B\left(R_{\rm o}-R_{\rm i}\right), 
\end{equation}
where we have simply rewrite equality $2r_c(eU_{\rm on})=R_{\rm o}-R_{\rm i}$
neglecting the evanescent modes.

When looking on the conductance spectra illustrated in
Fig.\ \ref{gmifma4pan:fig}(a) we see, for $B>0$, a~wide range of lower 
$U_{\rm eff}$ for which $G\approx{}0$, attached (via a~cusp region) to
the range of (approximately linearly) increasing $G$.
In order to determine the value of $U_{\rm on}(B)$  directly from the
conductance spectra $G(U_{\rm eff})$, we  find numerically the
value of $U_{\rm on}^{(1)}$ such that $G(U_{\rm on}^{(1)})=g_0$, and
$U_{\rm on}^{(2)}$ such that $G(U_{\rm on}^{(2)})=2g_0$, 
see the datapoints in Fig.\ \ref{gmifma4pan:fig}(a).  
Then, the linear extrapolation is performed to obtain
\begin{align}
  U_{\rm on}^{(0)} &= U_{\rm on}^{(1)} - \left(U_{\rm on}^{(2)}-U_{\rm on}^{(1)}\right)
  \frac{G(U_{\rm on}^{(1)})}{G(U_{\rm on}^{(2)})-G(U_{\rm on}^{(1)})} \nonumber \\
  &= 2U_{\rm on}^{(1)}-U_{\rm on}^{(2)},
  \label{uon0scal}
\end{align}
such that $G(U_{\rm on}^{(0)})\approx{}0$.
The resulting values of $U_{\rm on}^{(i)}$, depicted in
Fig.\ \ref{gmifma4pan:fig}(c) [datapoints], stay close to 
$U_{{\rm on},{\rm incoh}}$ obtained from Eq.\ (\ref{uonicoh}) [dashed line]. 

Remarkably, the values of the Fano factor corresponding to
$U_{\rm eff}=U_{\rm on}^{(i)}$, $i=1,2,$ see Fig.\ \ref{gmifma4pan:fig}(b),  
are close to $F_{\rm incoh}(B\rightarrow{}B_{c,2}-)\approx{}0.56$.
Similar observation applies for all studied values of $B\leqslant{}0.5\,$T,
see Fig.\ \ref{gmifma4pan:fig}(d); a~typical deviation does not exceed
$5\%$.

\begin{figure*}[!t]
  \includegraphics[width=\linewidth]{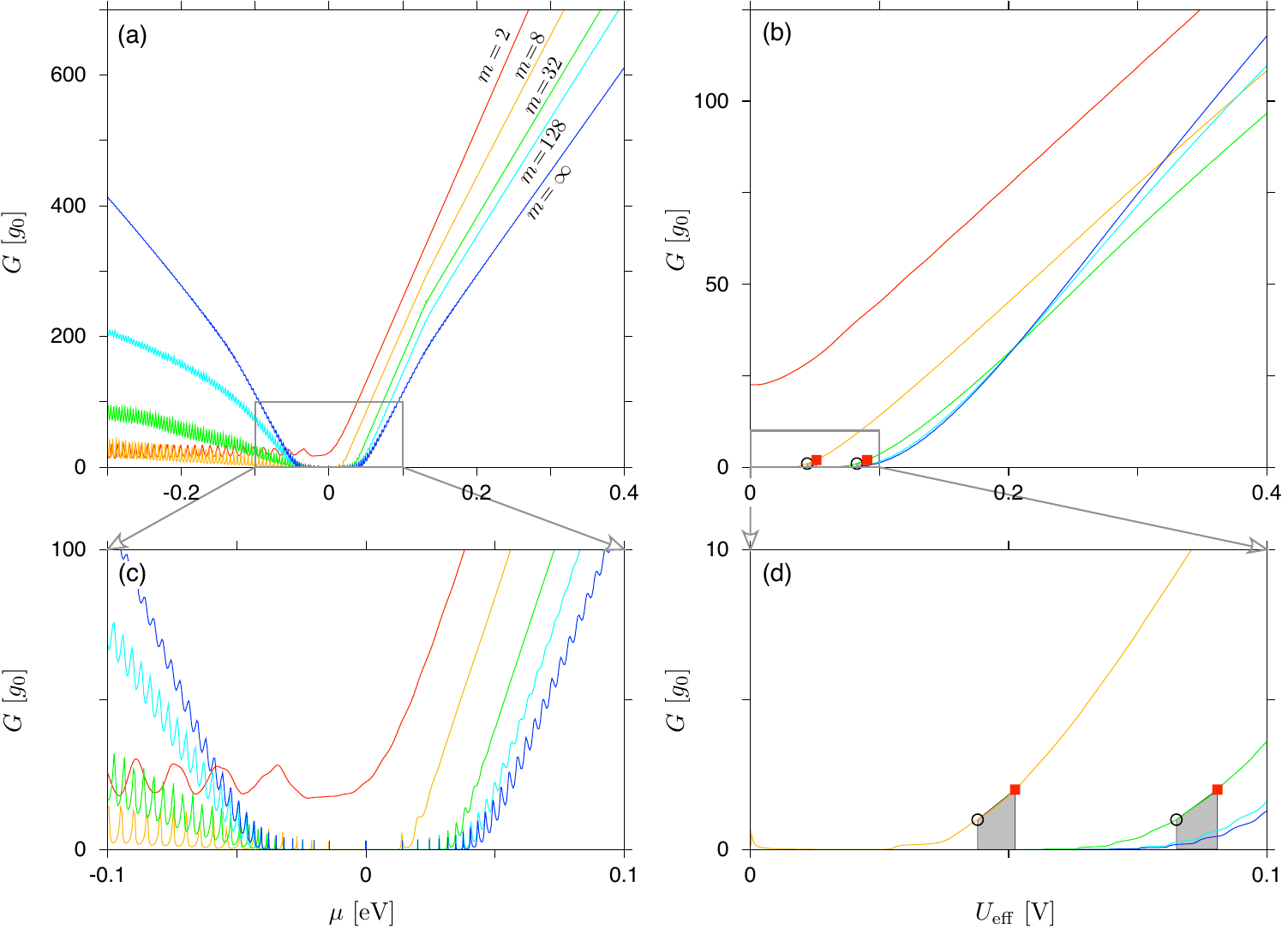}
  \caption{ \label{gfan4pan-mpots:fig}
  (a) Linear-response conductance as a~function of the chemical
  potential and (b) finite-voltage conductance, for $\mu=-eU_{\rm eff}/2$,
  as a~function of the voltage.
  The magnetic field is $B=0.2\,$T for all plots. 
  The disk radii are the same as in Fig.\ \ref{gfan4pan:fig}, but the
  barrier height, see Eq.\ (\ref{v0mpot})), is now fixed at
  $V_0=t_0/2=1.35\,$eV; the parameter $m$ is specified for each line.
  (c,d) Zoom-in, for low energies, with same datasets as in (a,b).
  Datapoints in (b,d) mark the values of $G(U_{\rm on}^{(i)})=ig_0$, $i=1,2,$ 
  defining the activation voltages $U_{\rm eff}=U_{\rm on}^{(i)}$. 
  }
\end{figure*}

\begin{figure*}[!t]
  \includegraphics[width=\linewidth]{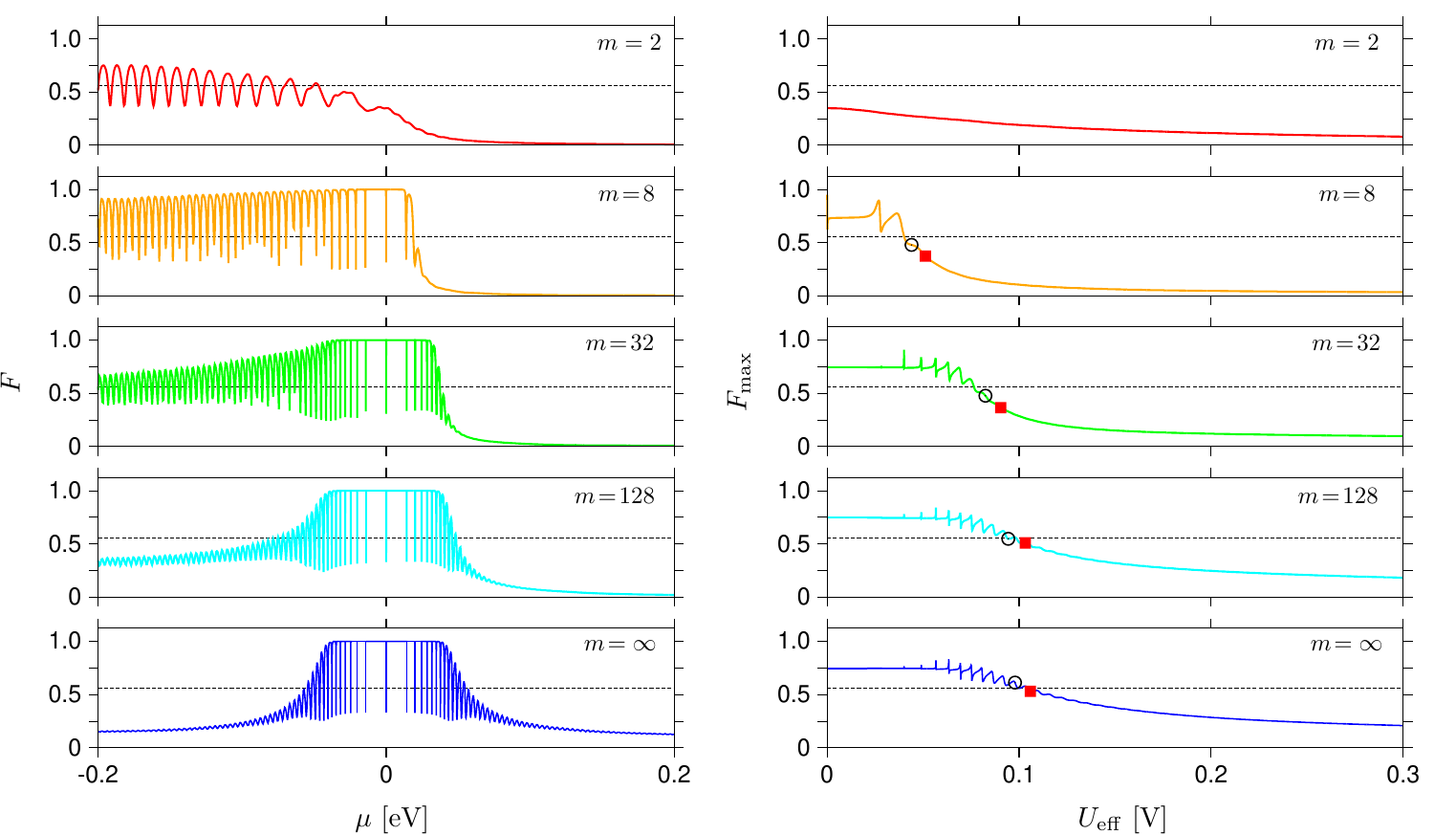}
  \caption{ \label{ffall10pan-mpots:fig}
  Left: Linear-response Fano factor as a~function of the chemical
  potential. Right: Finite-voltage Fano factor, for $\mu=-eU_{\rm eff}/2$,
  as a~function of the voltage.
  The magnetic field is $B=0.2\,$T for all plots, 
  the value of exponent $m$ is specified at each plot, and 
  remaining parameters are same as in Fig.\ \ref{gfan4pan-mpots:fig}.
  Horizontal line at each plot marks the value of
  $F_{\rm incoh}(B\rightarrow{}B_{c,2}-)=0.557898$, see Table~\ref{fanoriro}.
  Datapoints (right) mark the values of $F(U_{\rm on}^{(i)})$, $i=1,2,$ 
  corresponding to activation voltages $U_{\rm eff}=U_{\rm on}^{(i)}$, for
  which $G(U_{\rm on}^{(i)})=ig_0$ (see also Fig.\ \ref{gfan4pan-mpots:fig}). 
  }
\end{figure*}

\begin{figure*}[!t]
  \includegraphics[width=\linewidth]{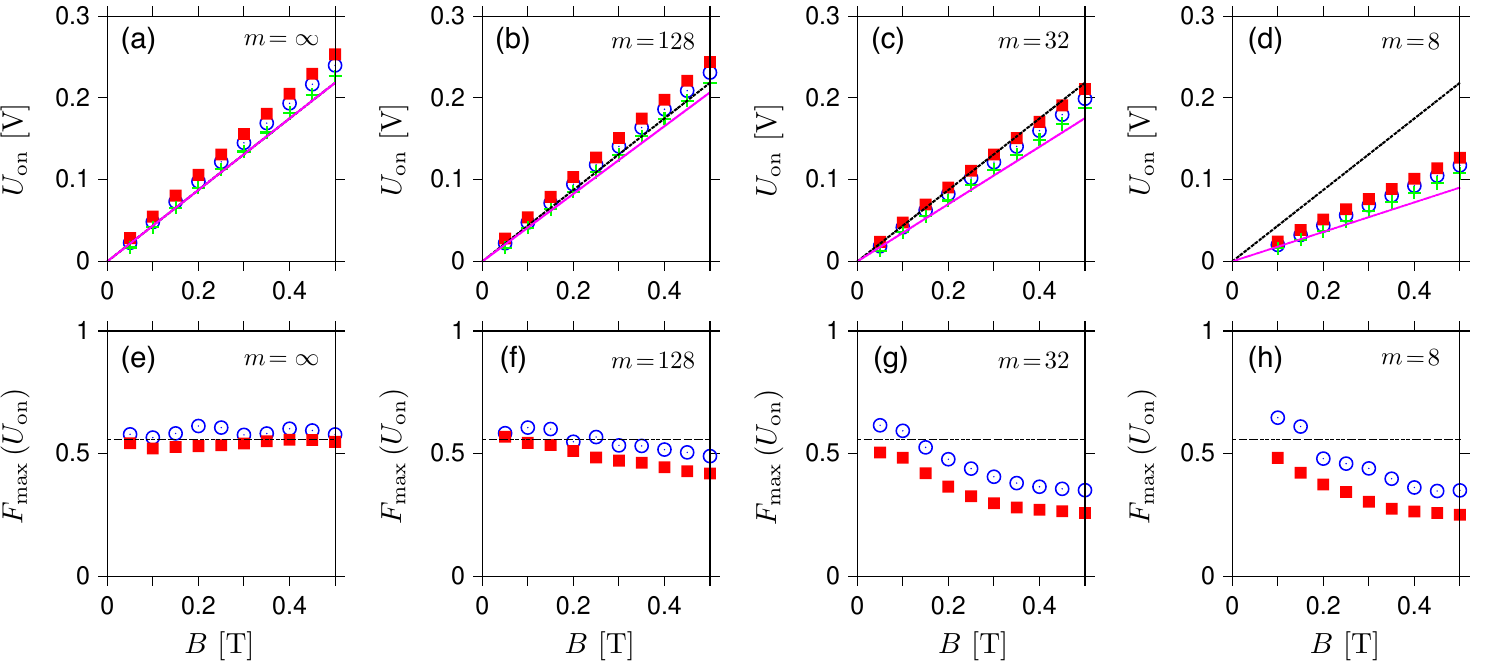}
  \caption{ \label{UonFma8pan-mpots:fig}
  (a--d) The activation voltage (for the definition, see Fig.\
  \ref{gmifma4pan:fig}) and (e--h) the corresponding Fano factor for
  $\mu=-eU_{\rm eff}/2$, displayed as functions of the magnetic field
  [datapoints]. 
  The value of exponent $m$ is specified at each plot;
  remaining parameters are same as in Fig.\ \ref{gfan4pan-mpots:fig}.
  Lines in (a--d) depict the approximation given by Eq.\ (\ref{uonicohdif})
  [purple solid] and Eq.\ (\ref{uonicoh}) [black dashed] coinciding in
  the $m\rightarrow{}\infty$ limit. 
  Horizontal lines in (e--h) mark the value of 
  $F_{\rm incoh}(B\rightarrow{}B_{c,2}-)=0.557898$, see Table~\ref{fanoriro}.
  }
\end{figure*}

\subsection{Smooth potential barriers}
In this subsection, we extend our numerical analysis onto smooth potential
barriers, defined by choosing $2\leqslant{}m<\infty$ in Eq.\ (\ref{v0mpot}).
Moreover, the barrier height is now finite, i.e., $V_0=t_0/2=1.35\,$eV, being
not far from the results of some first-principles calculations for
graphene-metal structures \cite{Gio08,Cus17}.
According to our best knowledge such a~model, first proposed in Ref.\
\cite{Ryc21b}, seems to be the simplest providing qualitatively correct
description of the conductance-spectrum asymmetry observed in existing
experiments \cite{Pet14,Kam21,Kum22}, in which the conductance for $\mu<0$
is noticeably suppressed, comparing to the $\mu>0$ range, due to the
presence of two circular p-n junctions in the former case.
(Such a~feature is also correctly reproduced by a~simpler model assuming the
trapezoidal potential barrier \cite{Par21}, allowing a~fully analytic
treatment, but this approach produces an artificial conductance maximum
near $\mu=0$.) 

The conductance spectra for five selected values of $m$ are displayed in
Fig.\ \ref{gfan4pan-mpots:fig}, both for the linear-response regime
[see Figs.\ \ref{gfan4pan-mpots:fig}(a,c)] and beyond [Figs.\
\ref{gfan4pan-mpots:fig}(b,d)]. This time, we have limited out presentation
a~single value of magnetic field, i.e., $B=0.2\,$T. 
It must be notice that finite value of $V_0$ results in small, but visible
spectrum asymmetry also for $m=\infty$.

The finite-voltage results, $G(U_{\rm eff})$ at $\mu=-eU_{\rm eff}/2$, allows
as to determine the activation voltage, $U_{\rm on}(B)$, in a~similar manner
as for an infinite-barrier case (see previous subsection).
When attempting to apply the incoherent-scattering approximation to
smooth potentials, some modification is required for 
Eq.\ (\ref{uonicoh}), which now can be rewritten as 
\begin{equation}
  \label{uonicohdif}
  U_{{\rm on},{\rm incoh}} = v_F{}BL_{\rm diff}(m).   
\end{equation}
In the above, we have introduced the $m$-dependent effective sample length
 given by \cite{Ryc21b,Ryc22}
\begin{equation}
  \label{ldiffm}
  {L_{\rm diff}(m)} = |R_{\rm o}-R_{\rm i}|
  \left(\frac{\hbar{}v_F}{|R_{\rm o}-R_{\rm i}|V_0}\right)^{1/m},   
\end{equation}
which reduces to $L_{\rm diff}(\infty)=R_{\rm o}-R_{\rm i}$ for a~rectangular
barrier, and gives $L_{\rm diff}(m=2)\ll{}R_{\rm o}-R_{\rm i}$ for the parabolic
case. In brief, Eq.\ (\ref{ldiffm}) can be derived by imposing
$V(\pm{}L_{\rm diff}/2)=-E_{\rm diff}$, where $E_{\rm diff}$ denotes the value
of Fermi energy, above which Sharvin conductance overrules the
pseudodiffusive conductance, namely, 
\begin{equation}
\label{efdifdef}
  E_{\rm diff}=\frac{\hbar{}v_F}{R_{\rm o}-R_{\rm i}}\approx{}1\,\text{meV}\ \ \
  \text{for}\ \ \ R_{\rm o}-R_{\rm i}=500\,\text{nm}. 
\end{equation}

In Fig.\ \ref{ffall10pan-mpots:fig} we show the Fano factor, for same
five values of $m$ as previously used for the conductance (see Fig.\
\ref{gfan4pan-mpots:fig}), and $B=0.2\,$T, as a~function $\mu$ in the
linear-response limit ($U_{\rm eff}\rightarrow{}0$), as well as a~function
$U_{\rm eff}$ for $\mu=-U_{\rm eff}/2$ (see left or right side of Fig.\
\ref{ffall10pan-mpots:fig}, respectively).
Again, the aperiodic oscillations almost vanish when entering the
nonlinear response regime; in fact, the shape of $F_{\rm max}(U_{\rm eff})$
appears to be much less sensitive to the value of $m$ than the
linear-response $F(\mu)$.
Datapoints on the right side of Fig.\ \ref{ffall10pan-mpots:fig},
identifying the values of $F(U_{\rm on}^{(i)})$, $i=1,2,$ such that
$G(U_{\rm on}^{(i)})=ig_0$ (see Fig.\ \ref{gfan4pan-mpots:fig}), are
available starting from $m=8$ (although the deviation from
$F_{\rm incoh}(B\rightarrow{}B_{c,2}-)\approx{}0.56$ is significant in such
a~case), whereas strong asymmetry of $F(\mu)$ is visible up to $m=32$. 

The values of $U_{\rm on}^{(i)}$ and the corresponding $F(U_{\rm on}^{(i)})$,
for the magnetic fields up to $B\leqslant{}0.5\,$T, are displayed in Fig.\ 
\ref{UonFma8pan-mpots:fig}. It can be noticed that the voltages
$U_{\rm on}^{(i)}$, see datapoints in Figs.\ \ref{UonFma8pan-mpots:fig}(a--d), 
show relatively good agreement with the approximation given by Eq.\
(\ref{uonicohdif}) [purple solid lines]; in fact, significant deviation
from Eq.\ (\ref{uonicoh}) relevant for the rectangular barrier [black dashed
lines] can be noticed for $m=8$ only.
On the contrary, corresponding Fano factors $F(U_{\rm on}^{(i)})$,
see datapoints in Figs.\ \ref{UonFma8pan-mpots:fig}(e--h),
stay close to the value of $F_{\rm incoh}(B\rightarrow{}B_{c,2}-)\approx{}0.56$
only for $m=\infty$ and $m=128$, showing that the incoherent treatment of the
shot-noise power, which we put forward in Sec.\ \ref{appcondfa},
is applicable only if the potential profiles is close to
(but not necessarily perfectly matching) the rectangular shape.

\section{Conclusions}
\label{conclu}

We have put forward an analytic description of the shot-noise power
in graphene-based disks in high magnetic field and doping. 
Assuming the incoherent scattering of
Dirac fermions between two potential steps of an infinite height, both
characterized by {\em a~priori} nonzero transmission probability due
to the Klein tunneling, we find that vanishing conductance should be
accompanied by the Fano factor $F\approx{}0.56$, weakly-dependent
on the disk proportions.

Next, the results of analytic considerations are confronted with the
outcome of computer simulations, including both rectangular and smooth
shapes of the electrostatic potential barrier in the disk area. 
Calculating both linear-response and finite-voltage transport
cumulants, within the zero-temperature Landauer-B\"{u}ttiker formalism,
we point out
that the role of evanescent waves (earlier ignored in the analytic
approach) is significant in the linear-response regime, however, one
should able to detect the quasi-universal $F\approx{}0.56$ noise in
a~properly designed experiment going beyond the linear response regime.
To achieve this goal, the following procedure is proposed: First,
the activation voltage (for a~fixed magnetic field) needs to be determined,
by finding a~cusp position
on the conductance-versus-voltage plot, above which the conductance
grows fast with the voltage (the average chemical potential is controlled
by the gate such that the conductance is minimal for a~given voltage). 
Having the activation voltage determined, one measures the noise
for such a~voltage, expecting the Fano factor to be close to
$F\approx{}0.56$. 

We expect that the effect we describe should be observable in ultraclean
samples and sub-kelvin temperatures (such as in Ref.\ \cite{Zen19});
for higher temperatures, hydrodynamic
effects may noticeably alter the measurable quantities \cite{Kum22}.
Since the noise-related characteristics seem to be generally
more sensitive to the potential shape then the conductance
(or the thermoelectric properties earlier discussed in Ref.\ \cite{Ryc23}),
the experimental study following the scenario presented here may be
a~suitable way to check whether the flat-potential area of a~mesoscopic size
is present or not in a~given graphene-based structure.

\section*{Acknowledgments}
The main part of the work was supported by the National Science Centre
of Poland (NCN) via Grant No.\ 2014/14/E/ST3/00256 (SONATA BIS).
Computations were performed using the PL-Grid infrastructure.


\appendix

\section{Numerical mode matching for smooth potentials}
\label{appnumoma}

Here we summarize the numerical approach earlier presented in Ref.\
\cite{Ryc23}.

In a~typical situation, system of ordinary differential
equations for spinor components $(\chi_a,\chi_b)$, 
see Eqs.\ (\ref{phapri}) and (\ref{phbpri}), needs to be integrated
numerically for all $j$-s.
In order to reduce round-off errors that may occur in finite-precision
arithmetics due to exponentially growing (or decaying) solutions, one can
divide the full interval, $R_{\rm i}<r<R_{\rm o}$, into $M$ parts, bounded by
\begin{align}
\label{rcllp}
  R_{c}^{(l)}&=R_{\rm i}+l\frac{R_{\rm o}-R_{\rm i}}{M} < r < R_{c}^{(l+1)},
  \nonumber \\
  & \text{with }\ \ l=0,1,\dots,M-1. 
\end{align}
(In particular, $R_{c}^{(0)}=R_{\rm i}$ and $R_{c}^{(M)}=R_{\rm o}$.) 

The wave function in the disk area $\chi_j^{({\rm disk})}$ is now given by
a~series of functions $\left\{\chi_j^{(l)}\right\}$ for $M$
consecutive intervals given by Eq.\ (\ref{rcllp}).
For the $l$-th interval, 
\begin{equation}
\label{chijdisknum}
  \chi_j^{(l)}=A_j^{(l)}\chi_j^{(l),{\rm I}}+B_j^{(l)}\chi_j^{(l),{\rm II}}, 
\end{equation}
where $\chi_j^{(l),{\rm I}}$, $\chi_j^{(l),{\rm II}}$ are two linearly
independent solutions obtained by integrating
Eqs.\ (\ref{phapri}), (\ref{phbpri}) with two different initial conditions, 
$\left.\chi_j^{(l),{\rm I}}\right|_{r=R_{\rm i}^{(l)}}=(1,0)^T$ and  
$\left.\chi_j^{(l),{\rm II}}\right|_{r=R_{\rm i}^{(l)}}=(0,1)^T$. 
$A_j^{(l)}$ and $B_j^{(l)}$ are complex coefficients (to be determined later). 

In particular, for $R_{\rm o}=2R_{\rm i}=1000\,$nm and $B<0.5\,$T considered
in this paper, it is sufficient to set $M=20$ and employ a~standard
fourth-order Runge-Kutta (RK4) algorithm with a~spatial step of $0.5\,$pm.
(For such a~choice, the output numerical uncertainties of transmission
probabilities $T_j$ are smaller then $10^{-7}$.) 

The matching conditions for the $M+1$ interfaces at $r=R_{\rm i}$,
$r=R_{c}^{(1)}$, \dots, $r=R_{c}^{(M-1)}$, and $r=R_{\rm o}$, can now be written
as
\begin{align}
  \chi_j^{({\rm inner})}(R_{\rm i}) &= \chi_j^{(0)}(R_{\rm i}),
    \label{match-i} \\
  \chi_j^{(l)}(R_{c}^{(l+1)}) &= \chi_j^{(l+1)}(R_{c}^{(l+1)}),
  \ \ \ 
  l = 0,\dots,M-2,
  \label{match-l} \\
  \chi_j^{(M-1)}(R_{\rm o}) &= \chi_j^{({\rm outer})}(R_{\rm o}),
  \label{match-o} 
\end{align}
and are equivalent to the Cramer's system of $2(M+1)$ linear equations for
the unknowns 
$A_j^{(0)}$, $B_j^{(0)}$, \dots, $A_j^{(M-1)}$, $B_j^{(M-1)}$, $r_j$, and $t_j$.

Writing down the spinor components appearing in Eqs.\ (\ref{match-i}),
(\ref{match-l}), and (\ref{match-o}) explicitly,
we arrive to \begin{widetext}
$$
  \left[
  \begin{matrix}
    -\chi_{j,a}^{\rm out}(R_{\rm i})
    & \chi_{j,a}^{(0),{\rm I}}(R_{c}^{(0)})
    & \chi_{j,a}^{(0),{\rm II}}(R_{c}^{(0)})
    & & & &  \\
    -\chi_{j,b}^{\rm out}(R_{\rm i})
    & \chi_{j,b}^{(0),{\rm I}}(R_{c}^{(0)})
    & \chi_{j,b}^{(0),{\rm II}}(R_{c}^{(0)})
    & & & &  \\
    0 &  \chi_a^{(0),{\rm I}}(R_{c}^{(1)})
    & \chi_{j,a}^{(0),{\rm II}}(R_{c}^{(1)})
    &  & & &  \\
    0 &  \chi_b^{(0),{\rm I}}(R_{c}^{(1)})
    & \chi_{j,b}^{(0),{\rm II}}(R_{c}^{(1)})
    & \ddots & & &  \\
    &  &  & \ddots &
    \chi_{j,a}^{(\overline{M}),{\rm I}}(R_{c}^{(\overline{M})}) 
    & \chi_{j,a}^{(\overline{M}),{\rm II}}(R_{c}^{(\overline{M})})
    & 0 \\
    &   &  &  &
    \chi_{j,b}^{(\overline{M}),{\rm I}}(R_{c}^{(\overline{M})}) 
    & \chi_{j,b}^{(\overline{M}),{\rm II}}(R_{c}^{(\overline{M})})
    & 0 \\
    &  &  &  &
    \chi_a^{(\overline{M}),{\rm I}}(R_{c}^{({M})}) &
    \chi_{j,a}^{(\overline{M}),{\rm II}}(R_{c}^{({M})})
    & -\chi_{j,a}^{\rm in}(R_{\rm o}) \\
    &  &  &  &
    \chi_b^{(\overline{M}),{\rm I}}(R_{c}^{({M})})
    & \chi_{j,b}^{(\overline{M}),{\rm II}}(R_{c}^{({M})})
    & -\chi_{j,b}^{\rm in}(R_{\rm o}) \\
  \end{matrix}
  \right]
$$
\begin{equation}
  \label{lsysABrt}
  \times\ 
  \left[
    \begin{matrix}
      r_j \\ A_j^{(0)} \\ B_j^{(0)} \\ \vdots
      \\ A_j^{(M-1)} \\ B_j^{(M-1)} \\ t_j \\
    \end{matrix}
  \right]
  =
  \left[
    \begin{matrix}
      \,\chi_{j,a}^{\rm in}(R_{\rm i})\, \\
      \,\chi_{j,b}^{\rm in}(R_{\rm i})\, \\
      0 \\
      \vdots \\
      0 \\
    \end{matrix}
  \right], 
\end{equation}
\end{widetext}
where we have defined $\overline{M}=M-1$,
\begin{equation}
  \label{chijleads}
  \chi_j^{\rm in} \!= \left(
  \begin{array}{c}
    H_{j-1/2}^{(2)}(Kr) \\
    iH_{j+1/2}^{(2)}(Kr) \\
  \end{array}
  \right),
  \ \ \ 
  \chi_j^{\rm out} \!= \left(
  \begin{array}{c}
    H_{j-1/2}^{(1)}(Kr) \\
    iH_{j+1/2}^{(1)}(Kr) \\
  \end{array}
  \right). 
  \end{equation} 
For heavily-doped leads ($V_0\rightarrow\infty$) the wave functions
given by Eq.\ (\ref{chijleads}) simplify to 
\begin{equation}
  \label{chijasym} 
  \chi_j^{({\rm in})} =
    \frac{e^{iKr}}{\sqrt{r}}
    \left(\begin{array}{c} 1 \\ 1 \\ \end{array}\right), 
    \ \ \ \ 
    \chi_j^{({\rm out})} =
    \frac{e^{-iKr}}{\sqrt{r}}
    \left(\begin{array}{c} 1 \\ -1 \\ \end{array}\right),
\end{equation}
with $K=|E+{}V_0|/(\hbar{}v_F)\rightarrow{}\infty$. 

As the linear systems for different values of $j$-s are decoupled,
standard software packages can be used to find their solutions. 
We have chosen the double precision LAPACK routine {\tt zgesv}, see Ref.\ 
\cite{zgesv99}.




\begin{thebibliography}{}
\bibitem{wronfoo}
An important step in the derivation of Eq.\ (\ref{tjhank}) is the 
recognition of the Wronskian of Hankel functions, cf.\ 
Ref.\ \cite{Abram}, Chapter 9, Eq.\ 9.1.17.
\end{thebibliography}
\end{document}